\documentstyle[12pt,epsf]{article}
\begin{document}

\vspace*{4cm}

\begin{center}
{\large \bf Spectroscopy of doubly charmed baryons:\\ $\Xi_{cc}^{+}$ and
$\Xi_{cc}^{++}$. 
}\\
\vspace*{5mm}
S.S. Gershtein, V.V. Kiselev, A.K. Likhoded and A.I. Onishchenko\\
{\sf State Research Center of Russia "Institute for High Energy Physics"} \\
{\it Protvino, Moscow region, 142284 Russia}\\
Fax: +7-095-2302337\\
E-mail: likhoded@mx.ihep.su
\end{center}
\abstract{
Using the quark-diquark approximation in the framework of Buchm\" uller-Tye
potential model, we investigate the spectroscopy of doubly charmed baryons:
$\Xi_{cc}^{++}$ and $\Xi_{cc}^{+}$. Our results include the masses, parameters
of radial wave functions of states with the different excitations of both 
diquark and light quark-diquark system. We calculate the values of fine and
hyperfine splittings of these levels and discuss some new features, connected
to the identity of heavy quarks, in the dynamics of hadronic and radiative 
transitions between the states of these baryons.  
}\\

\newpage
\section{Introduction}
For experiments at the running and planned high-energy machines such as LHC,
B-factories (HERA-B, for example) and Tevatron with high luminosities,
there are some hopes and possibilities to observe and study new hadrons,
containing two heavy quarks, like the doubly charmed baryons $(ccq)$ (here and
throughout this paper, q denotes a light quark u or d).
 
Looking for such projects, it is important to have reliable theoretical
predictions as a guide to the experimental searches for these baryons. In our
previous papers we have already studied both the production mechanisms of
doubly charmed baryons on the different future facilities \cite{PL},
\cite{1} and the
total inclusive lifetimes of such systems \cite{2}. In this work we would like
to address the question on the spectroscopy for these baryons in the framework
of potential approach in the Buchm\" uller-Tye model. The $(QQ^{\prime }q)$
spectroscopy was also considered in some other models \cite{3,4,5,6}.

The baryons with two heavy quarks include the features of both the dynamics in
the $D$-meson with a fast moving light quark, surrounding a static $\bar
3$-color core, and the dynamics of heavy-heavy mesons ($J/\Psi$, $B_c$,
$\Upsilon$), with two heavy quarks sensitive to the QCD potential at short
distances. So, a rich spectrum is expected. First, there are excitations due to
the relative motion of two heavy quarks in the quark-diquark approximation,
which we use here. Some comments on the validity of this approximation will be
given later. Second, we consider also the excitations of the light quark or
combined excitations of both degrees of freedom. 

As we will show in this paper, some new features, related with
the identity of two heavy quarks, arise in the
dynamics of doubly charmed baryons (or doubly beauty baryons). The hadronic or
electromagnetic transition between the $2P\to 1S$ diquark levels is forbidden
in the absence of interaction of diquark with the light quark or without taking
into account nonperturbative effects. The qualitative picture of this effect
will be discussed below, and a rigorous quantitative solution of this problem
will be given in our
subsequent papers.

This work is organized as follows. Section 2 is devoted to the determination
of masses, values of radial wave functions (for $S$-levels) and their first
derivatives (for $P$-levels) at the origin for both the diquark and the system
of light quark-diquark. In Section 3 we evaluate the fine and hyperfine
splitting of different levels for the above systems. Section 4
contains some comments on the radiative and hadronic decays of doubly charmed
baryons. And, finally, Section 5 draws our conclusion.

\section{Mass spectrum of doubly charmed baryons.}
Investigating the baryon spectroscopy, one faces a three body problem in the
framework of nonrelativistic quantum mechanics. Its reduced hamiltonian has the
following expression:
\begin{equation}
H = \frac{p_x^2}{M}+\frac{p_y^2}{M}+v({\bf x},{\bf y}),
\end{equation}
where ${\bf x}$, ${\bf y}$ are Jacobi variables:
\begin{eqnarray}
{\bf x} &=& {\bf r}_2 - {\bf r}_1,\\
{\bf y} &=& (2{\bf r}_3 - {\bf r}_1 - {\bf r}_2 )\sqrt\frac{m}{2M+m},
\end{eqnarray}
where $M$ is the heavy quark mass, and $m$ is the light quark mass.

There are several methods for a solution of three-body 
Schr\" odinger equation \cite{5}. The first way is the variational methods,
where one expands the wave-function in terms of the eigenstates for a symmetric
harmonic oscillator, or gaussians, at different ranges in the ${\bf x}$
and ${\bf y}$ coordinates, or uses the hyperspherical formalism.  The other
methods are the quark-diquark approximation and Born-Oppenheimer approximation.
The former is used in our evaluation of doubly charmed baryon spectrum.
The reason is that the ground state of $(ccq)$ consists of a localized $(cc)$
cluster surrounded by the light quark q, with the average distance
$\langle r_{cc}\rangle $ much less than $\langle r_{cq}\rangle $. However,
when one accounts for the radial or orbital excitations of diquark, the average
separation between the heavy quarks increases, and the quark-diquark structure
is distroied. So, in this region, our results for the mass spectrum of these
baryons are quite crude. 

Next, a dramatic simplification of $(ccq)$ dynamics is obtained in the
Born-Oppenheimer or adiabatic approximation. Two heavy quarks have much less
velocity than that of the light quark. When they move, the light quark wave
function readjusts itself almost immediately to the state of minimal energy.
Therefore, the calculation can be done in two steps: for any given ${\bf x}$,
one computes the binding energy $\epsilon({\bf x})$, which is, then, used as an
effective potential governing the relative motion of heavy quarks. This was
done in \cite{3}: first, in the nonrelativistic potential model and, second, in
a variant of MIT bag model. From our point of view, this method is the most
suitable for the baryon spectroscopy. However, in this work, we will be
satisfied with the accuracy, given by the quark-diquark approximation. 
   
In present work, we use the QCD-motivated potential \cite{7} given by 
Buchm\" uller and Tye \cite{8}, which was justified on the $J/\Psi$ and
$\Upsilon$ spectra with the following values of parameters:
\begin{equation}
m_c = 1.486~GeV,\quad m_q = 0.385~GeV,
\end{equation}
where the light quark mass is obtained by the fitting of theoretical
predictions with the Buchm\" uller-Tye potential for the $D$-meson mass
to its experimental value. Of course the motion of light quark is 
relativistic inside the $D$ meson, as well as inside the baryons under
consideraton and so can not be treated in the framework of nonrelativistic
quantum mecanics. But we think, that considering its motion in both cases
in the same way, we can obtain good approximation for the values of mass
levels of our baryons. 

Estimating the mass spectrum of diquark excitations, one has to take into
account the factor of 1/2 for the potential due to the fact, that the heavy
quarks inside the diquark are in color antitriplet state. At distances 
$\sim 0.6-0.8$ fm we have an attraction of quarks inside the diquark, so
we assume that the shape of potential in this region related with their 
pairwise interaction is the same as for quark - antiquark interaction in 
heavy mesons. At larger distances we can not tell anything about it, so 
our estimates for higher-lying levels is quite rough. But for low-lying 
energy levels in this system the wave function is already zero in the 
region of large distances, so that our predictions for them can be trusted. 
Then, solving the Schr\" odinger equations for the diquark and light quark 
excitations, one finds the diquark and $\Xi_{cc}^{++}$, $\Xi_{cc}^{+}$-baryons
mass spectra and the characteristics of radial wave functions for both the 
diquark and the system of light quark-diquark $R_d(0)$, $R_l(0)$, $R_d^{\prime
}(0)$
and $R_l^{\prime }(0)$, shown in Tables 1, 2 and 3. 

\begin{table}[t]
\begin{center}
\begin{tabular}{|c|c|c|c|c|c|}
\hline
Diquark state  & Mass (GeV) &  $\langle r^2\rangle ^{1/2}$ (fm) & Diquark state
& Mass (Gev)
&
$\langle r^2\rangle ^{1/2}$ (fm) \\ 
\hline 
1S & 3.16 & 0.58 & 3P & 3.66 & 1.36\\
\hline 
2S & 3.50 & 1.12 & 4P & 3.90 & 1.86\\ 
\hline 
3S & 3.76 & 1.58 & 3D & 3.56 & 1.13\\
\hline 
2P & 3.39 & 0.88 & 4D & 3.80 & 1.59\\
\hline 
\end{tabular} 
\end{center} 
\caption{The $(cc)$-diquark spectrum: masses and mean-square radii.}
\end{table}

\begin{table}[b]
\begin{center}
\begin{tabular}{|c|c|c|c|}
\hline 
n (diquark)  & $R_{d(nS)}(0)(R_{d(nP)}^{\prime } (0))$  & n (diquark) & 
$R_{d(nS)}(0)(R_{d(nP)}^{\prime }(0))$  \\
\hline 
1S & 0.530 & 2S & -0.452  \\ 
\hline
2P & 0.128 & 3P & -0.158  \\
\hline 
\end{tabular}
\end{center}
\caption{The characteristics of diquark radial wave functions 
$R_{d(nS)}(0)$ (in $GeV^{3/2}$), $R_{d(nP)}^{\prime } (0)$ (in $GeV^{5/2}$).}
\end{table} 

\begin{table}[t]
\begin{center}
\begin{tabular}{|p{30mm}|c|p{20mm}|p{30mm}|c|p{20mm}|}
\hline
$n_d (diquark)$~- $n_l (light~quark)$  & Mass (GeV)  & 
$R_{l(nS)}(0)$ $(R_{l(nP)}^{\prime } (0))$ & $n_d (diquark)$~- 
$n_l (light~quark)$ & Mass (GeV) & $R_{l(nS)}(0)$ $(R_{l(nP)}^{\prime }(0))$ \\
\hline
1S 1S & 3.56 & 0.499 & 1S 2P & 4.03 & 0.118\\
\hline
2S 1S & 3.90 & 0.502 & 2S 2P & 4.36 & 0.119\\
\hline
3S 1S & 4.16 & 0.505 & 3S 2P & 4.62 & 0.121\\
\hline
2P 1S & 3.79 & 0.501 & 2P 2P & 4.25 & 0.119\\
\hline
3P 1S & 4.06 & 0.504 & 3P 2P & 4.52 & 0.119\\
\hline
3D 1S & 3.96 & 0.503 & 3D 2P & 4.42 & 0.117\\
\hline
\end{tabular}
\end{center}
\caption{The mass spectra and characteristics of light quark radial wave
functions in the doubly charmed baryons: $\Xi_{cc}^{++}$ and $\Xi_{cc}^{+}$
with the different excitations of diquark, $n_d$, and light quark-diquark
system, $n_l$.}
\end{table} 

In calculations we assume, that the threshold mass value of doubly charmed
baryons is determined by the hadronic decay into $\Lambda_c$-baryon and
$D$-meson, and, hence, it equals 4.26 GeV \cite{9}. The threshold for the
stability of diquark is estimated from the following result, stated for a heavy
quark-antiquark pair \cite{10}: if a heavy quark and a corresponding antiquark

are separated by a distance greater than 1.4-1.5 fm, then the most
energetically favorable and probable configuration results in a production of
light quark-antiquark pair, which leads to the fragmentation into a pair of
flavored heavy mesons.  So, we suppose that the same critial distance scale can
be used for the colored diquark system, which results in the fragmentation of
diquark to the heavy meson and heavy-light diquark.

\section{Spin-dependent splitting.}

In accordance with the results of refs. \cite{11,12} and \cite{SS}, 
we introduce
the additional term to the potential to take into account the spin-orbital and
spin-spin interactions, causing the splitting of $nL$-levels in both the
diquark and light quark-diquark system ($n$ is the principal quantum number,
$L$ is the orbital momentum). So, it has the form
\begin{eqnarray}
V_{SD}^{(d)}({\bf r}) &=& \frac{1}{2}\left(\frac{\bf L\cdot S}{2m_c^2}\right)
\left( -\frac{dV(r)}{rdr}+
\frac{8}{3}\alpha_s\frac{1}{r^3}\right)\nonumber \\
&& +\frac{2}{3}\alpha_s\frac{1}{m_c^2}\frac{\bf L\cdot S}{r^3}+\frac{4}{3}
\alpha_s\frac{1}{3m_c^2}{{\bf S}_{c1}\cdot {\bf S}_{c2}}[4\pi\delta({\bf r})]\\
&& +\frac{1}{3}\alpha_s\frac{1}{m_c^2}(-\frac{1}{4{\bf L}^2 -3}\times [
6({\bf L\cdot S})^2+3({\bf L\cdot S})-2{\bf L}^2{\bf S}^2])\frac{1}{r^3},
\nonumber     
\end{eqnarray}
for the diquark splitting, and
\begin{eqnarray}
V_{SD}^{(l)}({\bf r}) &=& \frac{1}{2}\left(\frac{\bf L\cdot S_d}{2m_c^2}
+ \frac{2\bf L\cdot S_l}{2m_l^2}\right)
\left( -\frac{dV(r)}{rdr}+
\frac{8}{3}\alpha_s\frac{1}{r^3}\right)\nonumber \\
&& +\frac{2}{3}\alpha_s\frac{1}{m_c m_l}\frac{(\bf L\cdot S_d + 
2L\cdot S_l)}{r^3}+ 
\frac{4}{3}\alpha_s\frac{1}{3m_c m_l}{({\bf S}_{d}+{\bf L}_d)\cdot {\bf S}_{l}}
[4\pi\delta({\bf r})]\\
&& +\frac{1}{3}\alpha_s\frac{1}{m_c m_l}(-\frac{1}{4{\bf L}^2 -3}\times [
6({\bf L\cdot S})^2+3({\bf L\cdot S})-2{\bf L}^2{\bf S}^2\nonumber \\
&& -6({\bf L\cdot S_d})^2-3({\bf L\cdot S_d})+2{\bf L}^2{\bf S_d}^2])
\frac{1}{r^3}, \nonumber    
\end{eqnarray}
for the light quark-diquark system, where $V(r)$ is the phenomenological
potential (Buchm\" uller-Tye (BT) potential in our case), $S_l$ and
$S_d$ are the light quark and diquark spins, respectively.  The first term
in both expressions takes into account the relativistic corrections to the
potential $V(r)$. The second, third and fourth terms are the relativistic
corrections coming from the account for the one gluon exchange between the
quarks. $\alpha_s$ is the effective constant of quark-gluon interactions
inside the baryons under consideration.

Expression (6) for the additional part of potential, causing the
splitting of levels in the light quark-diquark system is obtained from
the summing of the pair interactions, like (5), for the light quark with each
of the heavy quarks. We include also the correction, connected to the
interaction of internal diquark orbital momentum with the light quark spin. 
 
The value of $\alpha_s$ parameter in (5), (6) can be determined in the 
following way. The splitting of the $S$-wave heavy quarkonium 
$(Q_1\bar Q_2)$ is determined by the expression 
\begin{equation}
\Delta M(nS) = \frac{8}{9}\alpha_s\frac{1}{m_1m_2}|R_{nS}(0)|^2,
\end{equation}
where $R_{nS}(0)$ is the radial wave function of the quarkonium, at the origin.
Using the experimental value of $1S$-state splitting in the $c\bar c$ system
\cite{13} 
\begin{equation} 
\Delta M(1S,c\bar c) = 117\pm 2 MeV 
\end{equation} 
and the $R_{1S}(0)$ value calculated in the potential model with the BT 
potential for the $c\bar c$ system, one gets the value of $\alpha_s(\Psi)$ 
coupling constant for the effective Coulomb interaction of heavy quarks.
 
In the present paper, we take into account the variation of the effective 
Coulomb coupling constant versus the reduced mass of the system $(\mu )$.
In the one-loop approximation at the momentum scale  $p^2$, the `running'
coupling constant in QCD is determined by the expression
\begin{equation} 
\alpha_s (p^2) = \frac{4\pi}{b\cdot\ln (p^2/\Lambda_{QCD}^2)}, 
\end{equation} 
where $b = 11 -2n_f/3$, and $n_f = 3$, when one takes into account the 
contribution by the virtual light quarks, $p^2< m_c^2$. We assume that
the average kinetic energies of $c$-quarks inside the diquark and the light
quark-diquark system (the average kinetic
energy weakly depends on the reduced mass of the system) are equal to:
\begin{equation}
\langle T_{d}\rangle \approx 0.2~GeV,
\end{equation}
\begin{equation}
\langle T_{l}\rangle \approx 0.4~GeV,
\end{equation}
so that, using the expression for the kinetic energy,
\begin{equation}
\langle T\rangle  = \frac{\langle p^2\rangle }{2\mu },
\end{equation}
where $\mu$ is the reduced mass of the system, one gets
\begin{equation}
\alpha_s(p^2) = \frac{4\pi}{b\cdot\ln (2\langle T\rangle \mu/\Lambda_{QCD}^2)},
\end{equation} 
so that $\alpha_s (\Psi) = 0.44$ at $\Lambda_{QCD}\approx 113~MeV$.

As one can see from equations (5) and (6), in contrast to the $LS$-coupling 
in the diquark, there is the $jj$-coupling in the light quark-diquark system,
where diquark and light quark have the different masses (here ${\bf LS}_l$ is
diagonal at the given ${\bf J}_l$ momentum,  $({\bf J}_l = {\bf L} + {\bf S}_l,
{\bf J} = {\bf J}_l + {\bf \bar J})$, $\bf J$ is the total spin of the system,
and $\bf\bar J$ is the total spin of the diquark (as we will see below in the
case of interest, $\bf\bar J$ equals to ${\bf S}_d$)).

To calculate the values of level shifts, appearing because of the spin-spin
and spin-orbital interactions, one has to average expressions (5), (6)
over the wave functions of the corresponding states. Then, because the leading
contribution to the spin-orbital splitting of the light quark-diquark
system is given by the term $\frac{1}{2}\frac{L\cdot
S_l}{2m_l^2}(-\frac{dV(r)}{rdr}+\frac{8}{3}\alpha_s
\frac{1}{r^3})$, we can use the state vectors  with the given values of $\bf J$
and ${\bf J}_l$ as the first approximation to the eigenvectors of the
potential. For the potential terms, which are not diagonal in these states, we
can choose another basis of vectors with the given values of $\bf J$ and
${\bf S} = {\bf S}_l + {\bf\bar J}$ 
\begin{equation}
|J;J_l\rangle  = \sum_{S} (-1)^{(\bar J+S_l+L+J)}\sqrt {(2S+1)(2J_l+1)}
\left\{\begin{array}{ccc} \bar J & S_l & S \\
                        L & J & J_l \end{array}\right\}|J;S\rangle 
\end{equation}
or ${\bf J}$ and ${\bf J}_d$
\begin{equation}
|J;J_l\rangle  = \sum_{Jd} (-1)^{(\bar J+S_l+L+J)}\sqrt {(2J_d+1)(2J_l+1)}
\left\{\begin{array}{ccc} \bar J & L & J_d \\
                        S_l & J & J_l \end{array}\right\}|J;Jd\rangle 
\end{equation}
so that the potential terms of the order of $1/m_cm_l$, $1/m_c^2$ lead,
generally speaking, to the mixing of levels with the different $J_l$ values at
the given $J$ values. The identity of heavy quarks results in $S_d=1$ at even
$L_d$ and $S_d=0$ at odd $L_d$, where $L_d$ is the diquark orbital momentum.
Table 3 shows us that we has to take into account only the spin-orbital
splitting of $1S 2P$ and $3D 1S$ levels.

In the first case (the splitting of the light quark-diquark system levels
$\Delta^{(J)}$ for $1S2P$) one has:
\begin{equation}
\Delta^{(\frac{5}{2})} = 17.4~MeV.
\end{equation}
The levels with $J = \frac{3}{2}$ (or $\frac{1}{2}$), but with the different
$J_l$, get the mixing. For $J =\frac{3}{2}$, the mixing matrix is equal to
\begin{equation}
\left(\begin{array}{cc} 4.3 & -1.7 \\
                        -1.7 & 7.8 \end{array}\right)~MeV 
\end{equation}
with the eigenvectors
\begin{eqnarray}
|1S2P(\frac{3}{2}^{\prime })\rangle  &=& 0.986|J_l=\frac{3}{2}\rangle
+0.164|J_l=\frac{1}{2}\rangle ,\\
|1S2P(\frac{3}{2})\rangle  &=& -0.164|J_l=\frac{3}{2}\rangle
+0.986|J_l=\frac{1}{2}\rangle ,
\nonumber
\end{eqnarray}
and the eigenvalues
\begin{eqnarray}
\lambda_1^{\prime } &=& 3.6~MeV,\\
\lambda_1 &=& 8.5~MeV.\nonumber
\end{eqnarray}
For $J=\frac{1}{2}$, the mixing matrix equals
\begin{equation}
\left(\begin{array}{cc} -3.6 & -55.0 \\
                        -55.0 & -73.0 \end{array}\right)~MeV 
\end{equation}
with the eigenvectors
\begin{eqnarray}
|1S2P(\frac{1}{2}^{\prime })\rangle  &=& 0.957|J_l=\frac{3}{2}\rangle
-0.291|J_l=\frac{1}{2}\rangle ,\\
|1S2P(\frac{1}{2})\rangle  &=& 0.291|J_l=\frac{3}{2}\rangle
+0.957|J_l=\frac{1}{2}\rangle ,
\nonumber
\end{eqnarray}
and the eigenvalues
\begin{eqnarray}
\lambda_2^{\prime } &=& 26.8~MeV,\\
\lambda_2 &=& -103.3~MeV.\nonumber
\end{eqnarray}
In the second case (the splitting of diquark levels
$\Delta^{(J_d)}$ for $3D1S$) one gets: 
\begin{eqnarray}
\Delta^{(3)} &=& -3.02~MeV,\nonumber\\
\Delta^{(2)} &=& 2.19~MeV,\\
\Delta^{(1)} &=& 3.39~MeV.\nonumber\\
\end{eqnarray}
For the spin-spin interactions inside the diquark, we have only a shift of the
energy levels because of the identity of heavy quarks.

The hyperfine splitting for the light quark-diquark system can be computed,
making use of the following formula:
\begin{equation}
\Delta_{h.f.}^{(l)} = \frac{2}{9}(S(S+1)-\bar J(\bar J +1 ) - \frac{3}{4})
\alpha_s(2\mu T)\frac{1}{m_cm_l}*|R_l(0)|^2,
\end{equation}
where $R_l(0)$ is the value of radial wave function in the system of light
quark-diquark at the origin, and
\begin{equation}
\Delta_{h.f.}^{(d) }= \frac{1}{9}
\alpha_s(2\mu T)\frac{1}{m_c^2}*|R_d(0)|^2,
\end{equation}
for the diquark level shifts, where $R_d(0)$ is the radial 
wave function of diquark at the origin.

For the $1S$ and $2S$-wave states of diquark
we have the following values of the shifts (because of the identity of
the quarks there are no splittings of these levels)
\begin{eqnarray}
\Delta (1S) &=& 6.3~MeV,\nonumber\\
\Delta (2S) &=& 4.6~MeV.\nonumber
\end{eqnarray}

The mass spectrum of doubly charmed baryons ($\Xi_{cc}^{++}$ and 
$\Xi_{cc}^{+}$), with the account for the calculated splittings
is shown in Fig.1 and Table 4.

\begin{figure}[t]
\setlength{\unitlength}{0.6mm}\thicklines
\begin{center}
\begin{picture}(240,200)
\put(15,56.6){\line(1,0){15}}
\put(15,59.6){$1S1S$}
\put(32,61){\line(1,0){15}}
\put(32,47.8){\line(1,0){15}}
\put(50,58){$^{3/2^{+}}$}
\put(50,44.8){$^{1/2^{+}}$}

\put(15,126){\line(1,0){15}}
\put(15,128){$1S2S$}

\put(62,148){\line(1,0){15}}
\put(62,151){$2P2S$}

\put(62,79){\line(1,0){15}}
\put(62,82){$2P1S$}
\put(79,83.4){\line(1,0){15}}
\put(79,70.2){\line(1,0){15}}
\put(97,80.4){$^{3/2^{-}}$}
\put(97,67.4){$^{1/2^{-}}$}

\put(15,90){\line(1,0){15}}
\put(15,93){$2S1S$}
\put(32,94.4){\line(1,0){15}}
\put(32,81.2){\line(1,0){15}}
\put(50,91.4){$^{3/2^{+}}$}
\put(50,78.2){$^{1/2^{+}}$}

\put(159,96){\line(1,0){15}}
\put(159,99){$3D1S$}
\put(179,94.7){\line(1,0){15}}
\put(179,96.2){\line(1,0){15}}
\put(179,97.34){\line(1,0){15}}
\put(200,108.9){\line(1,0){15}}
\put(220,106.59){$^{7/2^{+}}$}
\put(200,105){\line(1,0){15}}
\put(220,102){$^{5/2^{\prime +}}$}
\put(200,100.74){\line(1,0){15}}
\put(220,97.074){$^{3/2^{\prime +}}$}
\put(200,87.54){\line(1,0){15}}
\put(220,84.54){$^{1/2^{+}}$}
\put(200,83){\line(1,0){15}}
\put(220,80){$^{3/2^{+}}$}
\put(200,78.1){\line(1,0){15}}
\put(220,75.1){$^{5/2^{+}}$}

\put(107,103){\line(1,0){15}}
\put(107,106){$1S2P$}
\put(125,107.03){\line(1,0){15}}
\put(143,108.03){$^{1/2^{\prime -}}$}
\put(125,104.03){\line(1,0){15}}
\put(143,103.43){$^{5/2^{-}}$}
\put(125,103.15){\line(1,0){15}}
\put(143,99.045){$^{3/2^{-}}$}
\put(125,102.023){\line(1,0){15}}
\put(143,94.53){$^{3/2^{\prime -}}$}
\put(125,89.94){\line(1,0){15}}
\put(143,86.94){$^{1/2^{-}}$}

\put(15,116){\line(1,0){15}}
\put(15,119){$3S1S$}
\put(32,120.4){\line(1,0){15}}
\put(32,107.2){\line(1,0){15}}
\put(50,117.4){$^{3/2^{+}}$}
\put(50,104.2){$^{1/2^{+}}$}

\put(62,106){\line(1,0){15}}
\put(62,109){$3P1S$}
\put(79,110.4){\line(1,0){15}}
\put(79,97.2){\line(1,0){15}}
\put(97,107.4){$^{3/2^{-}}$}
\put(97,94.2){$^{1/2^{-}}$}

\put(10,115){\line(1,0){230}}
\put(185,118){$\Lambda_c~D$~~threshold}

\put(107,136.1){\line(1,0){15}}
\put(107,139.1){$2S2P$}

\put(107,162){\line(1,0){15}}
\put(107,165){$3S2P$}

\put(159,126.2){\line(1,0){15}}
\put(159,129.2){$2P2P$}

\put(199,142){\line(1,0){15}}
\put(199,145){$3D2P$}

\put(159,152){\line(1,0){15}}
\put(159,155){$3P2P$}

\put(10,0){\framebox(230,200)}
\put(0,0){$3.0$}
\put(10,50){\line(1,0){3}}
\put(0,50){$3.5$}
\put(10,100){\line(1,0){3}} 
\put(0,100){$4.0$}
\put(10,150){\line(1,0){3}}
\put(0,150){$4.5$}
\put(10,200){\line(1,0){3}}
\put(0,200){$5.0$}
\end{picture}
\end{center}
\caption{The spectrum of doubly charmed baryons: $\Xi_{cc}^{++}$ and 
$\Xi_{cc}^{+}$.}
\label{pic1}
\end{figure}
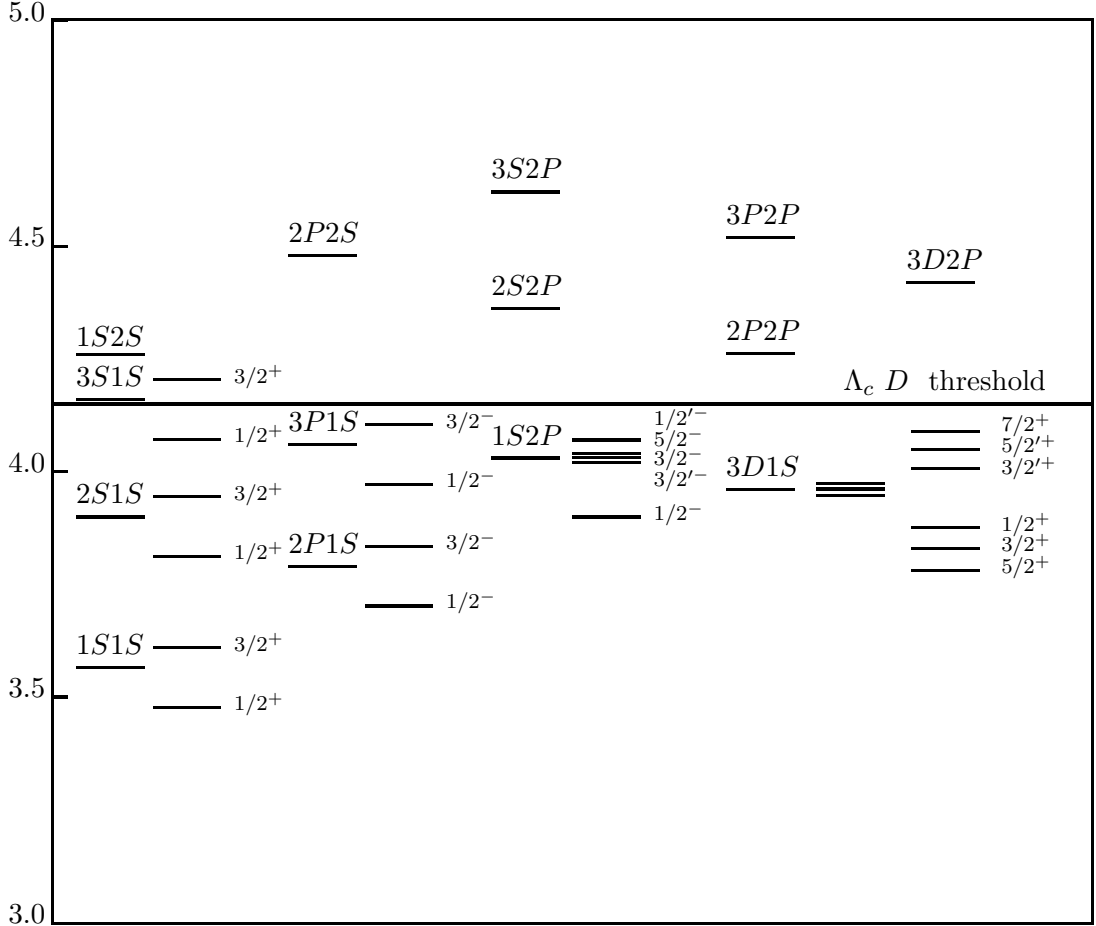

\begin{table}[t]
\begin{center}
\begin{tabular}{|p{40mm}|c|p{40mm}|c|}
\hline
$(n_d (diquark)$~- $n_l (light~quark))$, J$^{P}$ & Mass (GeV)
 & ($n_d (diquark)$~- 
$n_l (light~quark))$, J$^{P}$ & Mass (GeV)  \\
\hline
(1S 1S)$1/2^{+}$ & 3.478 & (3P 1S)$1/2^{-}$    & 3.972 \\
\hline
(1S 1S)$3/2^{+}$ & 3.61  & (3D 1S)$3/2^{\prime +}$ & 4.007 \\
\hline
(2P 1S)$1/2^{-}$ & 3.702  & (1S 2P)$3/2^{\prime -}$ & 4.034  \\
\hline
(3D 1S)$5/2^{+}$ & 3.781  & (1S 2P)$3/2^{-}$ & 4.039 \\
\hline
(2S 1S)$1/2^{+}$ & 3.812 & (1S 2P)$5/2^{-}$  & 4.047\\
\hline
(3D 1S)$3/2^{+}$ & 3.83 & (3D 1S)$5/2^{\prime +}$ & 4.05 \\
\hline
(2P 1S)$3/2^{-}$ & 3.834 & (1S 2P)$1/2^{\prime -}$ & 4.052 \\
\hline
(3D 1S)$1/2^{+}$ & 3.875 & (3S 1S)$1/2^{+}$   & 4.072\\
\hline
(1S 2P)$1/2^{-}$ & 3.927  & (3D 1S)$7/2^{+}$   & 4.089 \\
\hline
(2S 1S)$3/2^{+}$ & 3.944 & (3P 1S)$3/2^{-}$   & 4.104 \\
\hline
\end{tabular}
\end{center}
\caption{The mass spectra and characteristics of light quark radial wave
functions of doubly charmed baryons: $\Xi_{cc}^{++}$ and $\Xi_{cc}^{+}$ with
the different excitations of diquark, $n_d$, and light quark-diquark system,
$n_l$.}
\end{table}

\section{Transition between diquark levels: $2P\to 1S$ -- a laboratory
for the long distance QCD?}

Our plans for a future include a detail investigation of hadronic and radiative
transitions in the spectrum of doubly charmed baryons. However, here we would
like to discuss some new futures in the dynamics of baryons, containing
two identical heavy quarks.

So, because of the identity of two heavy quarks, we have a 
metastable $2P$-wave diquark state. This state has the $L=1$, $S=0$ quantum
numbers, and, hence, the transition to the ground state $(L=0, S=1)$ would
require the simultaneous change of orbital and spin quantum numbers. It is
worth to stress that the existence of such state is possible only in the
baryons with two heavy quarks, because in ordinary heavy baryons the light
diquark is never excited because of extremely large size. We have two possible
scenario for the realization of such transition:

1. A three-particle interaction via three-gluon vertex. This interaction breaks
down quark-diquark picture and leads to some new wave functions of these states
in the form: $C_1|L=1,S=0\rangle  + C_2|L=0,S=1\rangle $, where $|C_2|\ll
|C_1|$ for $2P$. So, as one can easily see, the account for
such interactions makes the radiative $M1$-transition to the ground state to be
possible.

2. A nonperturbative transition given by an operator, proportional to $\mu_B\;
\vec r\cdot\vec\nabla \;\vec H(0)\cdot(\vec S_1 - \vec S_2)$, where $\mu_B$ is 
the Bohr magneton, $\vec H(\vec r)$ is the chromomagnetic field and $\vec r$ is
the distance between two heavy quarks (here we have the same
situation as in the case of transition between the orto- and para-hydrogen).
This transition goes with the emission of pion. We would like to stress the
nonperturbative nature of this transition, because for its realization, the
necessary condition is that the chromomagnetic field at the different points is
correlated. The absence of such correlation prevents this scenario of
transition from realization, because it would require two consequent gluon
exchanges with the light quark or quark-gluon sea, wherein the orbital or spin
quantum numbers of state would change, what is impossible because of the
identity of two heavy quarks.
 
A detail quantitative investigation of the nature for this transition is
a subject of our subsequent papers. It will allow us to gain more information
on the nonperturbative dynamics of QCD (in the case of second type transition)
and, particularly, on the behaviour of chromomagnetic field at large distances.
So, we close this section with the question, posed in its title.
   
\section{Conclusion}

In this work we have used the Buchm\" uller-Tye potential model within the
quark-diquark approximation to describe the mass spectrum of
doubly charmed baryons: $\Xi_{cc}^{++}$ and $\Xi_{cc}^{+}$, including
the fine and hyperfine splittings of mass levels. We have discussed the
uncertainties involved and some possible ways to reduce them: the use of
Born-Oppenheimer approximation and the corrected potential for the level
splitting. In the previous section of article we have considered a new
phenomena, arising in the radiative or hadronic transitions of $2P\to 1S$
diquark levels. We have commented on some possible scenaria of this transition
and lessons, we could learn from a studying the nature of such transitions.

The authors express their gratitude to Prof. A.Wagner and DESY Theory Group for
a kind hospitality duiring a visit, when this paper was produced.
This work is in part supported by the Russian Foundation for Basic Research,
grants 96-02-18216 and 96-15-96575. A.I. Onishchenko aknowledges the support by
the NORDITA grant.

\end{document}